\documentstyle[epsfig]{aipproc}

\begin{document}
\title{A GRB Tool Shed}

\author{David J. Haglin$^{\dagger}$ 
  Richard J. Roiger$^{\dagger}$, Jon Hakkila$^{\dagger}$ \\
  Geoffrey Pendleton$^{\ddag}$, Robert Mallozzi$^{\ddag}$
}

\address{%
   $^{\dagger}$Minnesota State University, Mankato, MN 56001 \\
   $^{\ddag}$University of Alabama, Huntsville, AL 35812
}

\maketitle

\begin{abstract}
We describe the design of a suite of software tools to allow users 
to query Gamma Ray Burst (GRB) data and perform data mining 
expeditions.  We call this suite of tools a shed ({\bf SH}ell for
{\bf E}xpeditions using {\bf D}atamining).  
Our schedule is to have a completed
prototype (funded via the NASA AISRP) by February, 2002.
Meanwhile, interested users will find a partially functioning tool shed
at {\tt http://grb.mankato.msus.edu}.
\end{abstract}

\section*{Introduction}

We are implementing a suite of software tools to aid Gamma-Ray
Burst (GRB) researchers in working with the GRB data.  The major
features of the tool shed --- a {\bf SH}ell for {\bf E}xpeditions using 
{\bf D}atamining --- are a web-based data query facility, web-based
data visualization capability, and a web-based interface to 
data mining software tools.
The tool shed maintains a database of users allowing each user 
to store their own work at the tool shed site.  
Each user's data will not be visible to other users of the system.

Our GRB tool shed is populated with a standard set of preprocessed GRB 
data such as the basic table data, flux/fluence data, and 
duration \cite{4BCAT}.
These data are stored in tabular form as rows (burst instances) and 
columns (attributes).  Each burst has the same attributes as all other
bursts, with a provision for indicating ``missing'' attributes.  Users
may augment this database by uploading their own table or performing
SQL database queries for data selection and calculations.

The data mining tools can be given any of the queried data and produce 
either rules for classification of bursts or an identification of 
classes of bursts by identifying which bursts belong to which class.
Note that these ``identified'' classes are not necessarily classes based 
on physical properties of bursts; they may be due to instrumental 
bias, or even statistical fluctuations from the small numbers of 
instances (c.f. \cite{HHP+99}).
This step can lead to defining new data in new tables for 
further exploration.
An example of applying these tools is given in \cite{HHR+99}.

\subsection*{Data Storage}

Once a registered user has logged in, the data can be manipulated in a
variety of ways.
The user may simply query the existing data, selecting a subset of the 
bursts and/or attributes for later processing.
They may augment the data by uploading their own attributes for 
existing bursts.
Or they may augment the data by uploading information for new bursts.

As an example database query, consider selecting all bursts from 
the 4B catalog with high relative measurement errors on the channel 1 
fluence measurement.
To determine what might be considered ``high'', it may be necessary
to see all 
of the bursts sorted by relative measurement error, easily done with 
this SQL query:
\begin{quote}\tt
SELECT burstnum, channel1FluenceError FROM 4BFluxTable \newline
WHERE channel1Fluence <> 0 \newline
ORDER BY (channel1FluenceError/channel1Fluence);
\end{quote}

Now, after viewing the results of this query, one may decide that 
a ``quality'' threshold of 1.0 standard deviations would be required
on the fluence data.
The quality data can be extracted using this query:
\begin{quote}\tt
SELECT burstnum, channel1Fluence, channel1FluenceError \newline
FROM 4BFluxTable WHERE channel1Fluence <> 0 \newline
\quad AND (channel1FluenceError/channel1Fluence) > 1.0;
\end{quote}

For those users  
unfamiliar with SQL,
the web application will provide point-and-click, 
fill-in-the-box forms for generating a database query.
These web pages should provide the user with most of the features they would 
be interested in.
There will also be a blank form for users who may wish to enter the SQL 
query directly.
Either way, the names of the tables and attributes will be shown on the 
web browser so the user need not memorize them.

\subsection*{Data Visualization}

At any time the user may decide to invoke visualization tools to 
help ``see'' the data in their scratch area or in the system tables.
The web application will send these requests to ION (IDL On the Net) 
to produce graphical views of the data.
The display of these graphs will be done on the user's web browser window.

\subsection*{Data Mining Tools}

Several data mining tools will be available as part of the web application.
Initially, there will be the classic classification tool, C4.5 \cite{C45},
a tool developed by a member of our group, ESX \cite{RGHH99}, 
and at least one Neural Network package.
These software packages run on the web server and the web application 
will guide the user, requesting information needed by the specific
data mining tool being invoked.

The output from the data mining tools differ from one tool to the next.
Our web application will be able to capture the output and transform
it to an internal ``rule'' format.
Once captured, the user may view the rules, 
or go even one step further by applying the 
rules to a database in their scratch area.
The application of the rules to the data is a significant feedback 
component of the tool shed.

\subsection*{Online Help}

There will be extensive online help to guide the user through the data 
mining process.  
The help system will be written in a hyper-linked book format complete with 
a table of contents, an index, and a search engine.  
There will also be context-sensitive help in the sense of hyper-links from 
web forms to relevant pages of the help system.  
The help system will address how to use the tool.   
The GRB data will be minimally documented, with pointers (hyper-links) to 
existing help in understanding the meaning of the data attributes.

A tutorial will be provided that brings the user through a data mining 
session.  
This tutorial will provide (scientific) rationale for 
selecting options along with way.  
Although the tool shed need will not be specific to GRB data, the 
tutorial will be.

\section*{GRB Tool Shed System Platform}

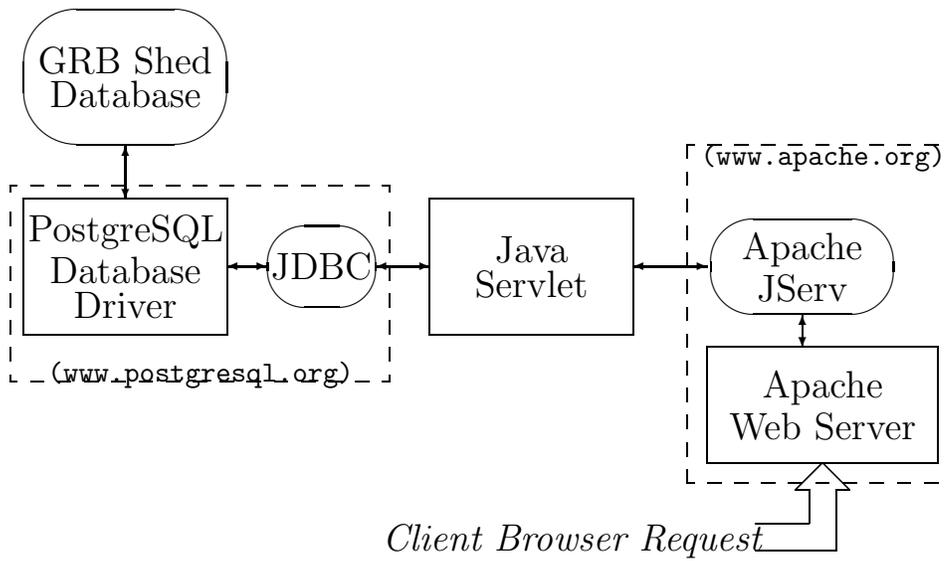
\begin{figure}[ht!] 
\begin{center}\large
\unitlength=0.9mm
\begin{picture}(140,80)
\put(55,0){\it Client Browser Request}
\put(110,0){\line(1,0){12}}
\put(110,4){\line(1,0){8}}
\put(118,4){\line(0,1){5}}
\put(122,0){\line(0,1){9}}
\put(122,9){\line(1,0){2}}
\put(118,9){\line(-1,0){2}}
\put(116,9){\line(1,1){4}}
\put(124,9){\line(-1,1){4}}
\put(100,10){\dashbox{2}(40,50)[ct]{\small\tt (www.apache.org)}}
\put(103,13){\framebox(34,17)[cc]{\shortstack{Apache \\ Web Server}}}
\put(117,42){\oval(27,14)}
\put(117,42){\makebox(0,0){\shortstack{Apache \\JServ}}}
\put(117,30){\vector(0,1){5}}
\put(117,35){\vector(0,-1){5}}
\put(62,32){\framebox(30,20)[cc]{\shortstack{Java \\Servlet}}}
\put(103,42){\vector(-1,0){11}}
\put(92,42){\vector(+1,0){11}}
\put(0,25){\dashbox{2}(56,29)[cb]{\small\tt (www.postgresql.org)}}
\put(46,42){\oval(16,12)}
\put(46,42){\makebox(0,0){JDBC}}
\put(62,42){\vector(-1,0){8}}
\put(54,42){\vector(+1,0){8}}
\put(2,32){\framebox(30,20)[cc]{\shortstack{PostgreSQL \\ Database \\ Driver}}}
\put(38,42){\vector(-1,0){6}}
\put(32,42){\vector(+1,0){6}}
\put(17,52){\vector(0,1){8}}
\put(17,60){\vector(0,-1){8}}
\put(17,70){\oval(30,20)}
\put(17,70){\makebox(0,0){\shortstack{GRB Shed \\ Database}}}

\end{picture}
\end{center}
\caption{Information Flow}
\label{fig:infoFlow}
\end{figure}

The GRB tool shed will initially run on a pentium-class computer system 
running RedHat Linux.
To maximize portability, the Java language will be used.  
Since this tool is web-based, all of the Java code will run as Java Servlets.  
The Apache web server will be used along with its companion Java Servlet 
Apache-JServ.  
The GRB data will be stored in a PostgreSQL database with access to the 
Java code provided via JDBC.  
All of these software packages are available on many platforms, including 
Unix, Windows, and Mac, and they are all freely available.  
The flow of information through these packages is shown in 
Figure~\ref{fig:infoFlow}.

This whole process is initiated by aiming a web browser at the 
appropriate URL ({\tt http://grb.mankato.msus.edu/}) where the Apache 
web server is configured to start a Java Servlet via Apach JServ.  
The Java Servlet will then 
make 
requests to the PostgreSQL database via the JDBC package that comes with 
PostgreSQL.  
The major development effort for this application is in the creation 
of the Java Servlet code, which we call ``the web application.''

\section*{GRB Tool Shed Data Flow}

The Java Servlet application is a complex set of Java code with many data 
structures and interfaces.  
There are two major data formats that will be used: Standard User Interface 
Format (SUIF) and Standard Internal Classifier Format (SICF).  
The SUIF will be used when presenting data to the user on the client 
machine (web  browser).  
And SICF will be used when presenting data to any of the supported 
classifiers (data mining tools).  
The SICF format is proposed as a tool-independent representation format 
for holding all information necessary to conduct a 
classification/data mining run. 

We expect to provide a user interface on the web browser that looks very 
much like a spreadsheet program.  
This familiar view will allow the user to inspect and possibly update 
data easily.  

\begin{figure}[htbp]
\centerline{\epsfig{file=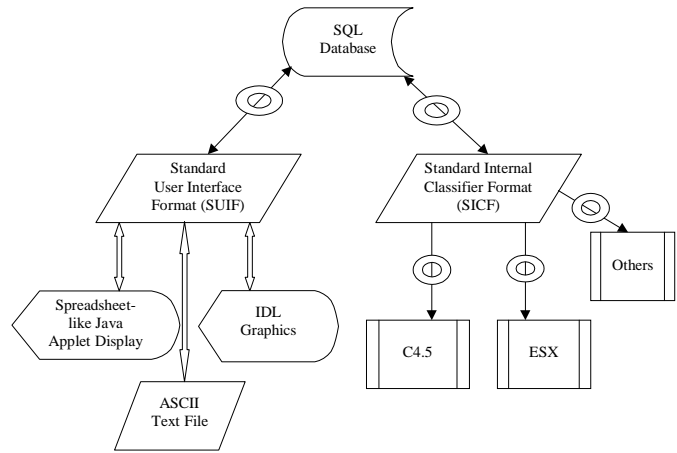,width=3.5in}}
\vspace{10pt}
\caption{Data Flow}
\label{fig:dataFlow}
\end{figure}

Note that a ``donut'' object indicates a data format conversion is needed at
that point.
Some of these data format conversions may require additional information, 
which would require the use of web pages to interact with the user.  
And the wide arrows indicate data flowing across the internet from server 
to client (browser). 

The control of when and where the data flows is completely independent of 
this diagram.  
Imagine a controller sitting above this page directing the data to 
flow along the various paths of the diagram.  
This controller is a web-based menu system under the direction of the user.

\section*{Web Pages (Control Flow)}

The design of this perspective of the GRB Tool Shed is least developed 
at this time.  
It is clear there needs to be a login menu screen.  
Once that information has been verified, the user will be placed in 
a main menu/dispatcher web page.  
This page will allow the user to click on various functions on a 
menu bar either along the side, the top, or the bottom.  
The menu selections/actions might be:

\begin{tabular}{@{\hspace*{4mm}}ll}
$\bullet$ graph/visualize data & 
  $\bullet$ show data in spreadsheet form \\
$\bullet$ initiate upload/download of data & 
  $\bullet$ work with the data (query database) \\
$\bullet$ initiate classification &
  $\bullet$ view documentation \\
\end{tabular}

\smallskip

Each of these items constitutes a large implementation effort.  
For example, the ``work with the data'' selection initiates a sequence of 
web pages/forms where the user is required to make selection criteria for the 
various rows and columns of the data.   
The ``initiate classification'' selection will cause the user to be 
prompted for which classifier tool to invoke, then guide the user 
through selecting parameter values that are specific to that tool.

\section*{Conclusions}

With the creation of this software comes a powerful research tool 
capable of automating many aspects of manipulating GRB data.  
Our goal is to go beyond the tool creation and build up the 
collection of attributes about the GRBs, emerging as a significant 
repository of GRB information with a built-in efficient methodology.

Our implementation strategy is to incrementally develop components.  
That way, as the development progresses, the web application will contain 
some usable software and data.

\end{document}